\documentclass[sigconf]{acmart}
\acmConference[Conference acronym 'XX]{Make sure to enter the correct
  conference title from your rights confirmation emai}{June 03--05,
  2018}{Woodstock, NY}
\usepackage{xcolor}
\usepackage{tcolorbox}
\newcommand{\blank}[1]{\hspace*{#1}}

\AtBeginDocument{%
  \providecommand\BibTeX{{%
    \normalfont B\kern-0.5em{\scshape i\kern-0.25em b}\kern-0.8em\TeX}}}

\setcopyright{acmcopyright}
\copyrightyear{2024}
\acmYear{2024}
\acmDOI{XXXXXXX.XXXXXXX}

\begin{document}

\title{Large language models for generating rules, yay or nay?}

\author{Shangeetha Sivasothy, Scott Barnett, Rena Logothetis, Mohamed Abdelrazek, Zafaryab Rasool, Srikanth Thudumu, Zac Brannelly}
\affiliation{%
\institution{Deakin University}
\city{Geelong}
\country{Australia}
}
\email{{s.sivasothy, scott.barnett, rena.logothetis, mohamed.abdelrazek, zafaryab.rasool, srikanth.thudumu, zac.brannelly}@deakin.edu.au}

\renewcommand{\shortauthors}{Sivasothy et al.}

\begin{abstract}
Engineering safety-critical systems such as medical devices and digital health intervention systems is complex, where long-term engagement with subject-matter experts (SMEs) is needed to capture the systems' expected behaviour. In this paper, we present a novel approach that leverages Large Language Models (LLMs), such as GPT-3.5 and GPT-4, as a potential world model to accelerate the engineering of software systems. This approach involves using LLMs to generate logic rules, which can then be reviewed and informed by SMEs before deployment. We evaluate our approach using a medical rule set, created from the pandemic intervention monitoring system in collaboration with medical professionals during COVID-19. Our experiments show that 1) LLMs have a world model that bootstraps implementation, 2) LLMs generated less number of rules compared to experts, and 3) LLMs do not have the capacity to generate thresholds for each rule. Our work shows how LLMs augment the requirements' elicitation process by providing access to a world model for domains.  
\end{abstract}



\maketitle

\section{Introduction}

Large Language Models (LLMs) have great potential to improve software engineering through human like generation of code and documentation. LLMs are also touted to include a "world model"~\cite{de2023chatgpt}, an understanding about our word learned from analysing internet scale datasets (including a model of every domain). Exploration of how the world model within an LLM can be exploited for software engineering tasks is an emerging area of research \cite{zhao2023survey, nejjar2023llms, chen2023effectiveness, du2023resolving}.  However, a central concern with using LLMs is their tendency to make up answers that sound plausible but are inaccurate \cite{hou2023large, ji2023survey, dhuliawala2023chain}. This suggests that LLMs are best suited to augmentation tasks prior to human review. One such task is the generation of rules pertaining to tacit knowledge of subject-matter experts.

In the health care domain, subject-matter experts specify logical rules for remotely monitoring patients. For new applications, the set of rules and thresholds is not immediately obvious and require time from clinicians to a) extract requirements \cite{liu2023radiology, zhou2023skingpt}, b) evaluate prognosis/diagnosis algorithms \cite{balas2023conversational, feng2023large, liu2023radiology, shyr2023identifying, zhang2023huatuogpt, zhou2023skingpt}, and c) validation in a clinical context \cite{hiesinger2023almanac, liu2023using, rao2023evaluating}. All this takes clinicians away from the important role of caring for patients. The goal of this study is to investigate if the world model in LLMs could reduce this time by acting as a proxy for subject-matter experts by scaffolding the initial rule sets. 

In this paper, we present a novel approach for software developers to collaborate with subject-matter experts on creating logical rules. We test the feasibility of our approach by conducting an experiment with four LLM prompting techniques (instruction following \cite{kojima2022large, prystawski2023think}, imitation \cite{shanahan2023role}, chain of thought \cite{wei2022chain}, and few-shot \cite{zhu2023large}) and two different LLMs (GPT-3.5 and GPT-4). The generated rules were compared to the rules from an industry case study, the Pandemic intervention Monitoring System (PiMS) where rules were specified manually by clinicians \cite{logothetis2022pims}. The benefits of our proposed approach include a) reducing implementation costs, and b) faster validation time of clinical rules (through rule and code synthesis). Our work is validated in the domain of remote health monitoring applications, with application to other domains with complex business logic. 

The contributions arising from this work include:
\begin{itemize}
    \item An initial empirical evaluation on the use of the world model in LLMs for the elicitation of tacit knowledge. 
    \item An approach to bootstrap logic rules during development of software systems for rapid validation of rules using LLMs.
    \item A comparison between LLMs generated rules and rules specified by subject-matter experts for an industry case study, PiMS. Evaluation includes the number of rules, average number of conditions in each rule, and the overlap between variables.
\end{itemize}

\begin{figure*}
   \includegraphics[width=\linewidth]{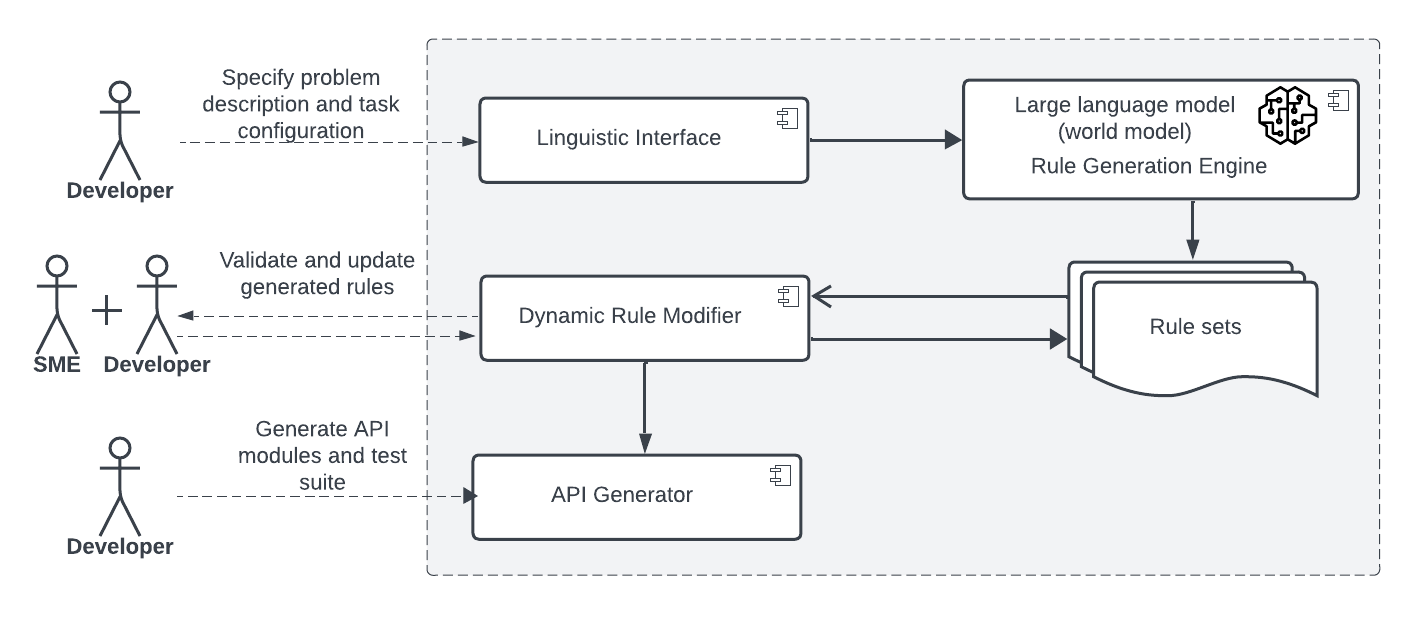}
   \caption{Overview of RuleFlex for generating and validation rules with subject-matter experts: 1) linguistic interface for specifying problem description, 2) rule generation engine for generating rule sets, 3) dynamic rule modifier for validating and updating the generated rule sets, and 4) API generator for production deployment.}  \label{fig:proposed_approach}
\end{figure*}

\textbf{Motivating Example:} Imagine Jack, a software engineer, involved in building a remote monitoring system for COVID-19 patients. He faces the common challenge of insufficient initial data due to privacy concerns \cite{lavin2022technology}. The remote monitoring system is a safety-critical system where a minor error in logic rules can result in a delay of clinical care that can be fatal. Jack can generate logic rules using a world model and can approach clinicians (i.e. SMEs) to validate the generated logic rules. But, clinicians do not have the time to work with Jack to evaluate and there does not exist universal evaluation between available logic rules in a world model and logic rules used by clinicians (i.e. SMEs). In addition, Jack is aware that these rule sets will need to be modified when developing a prototype for another remote monitoring system. Therefore, Jack wonders how to obtain the initial set of logic rules to develop a prototype, to view changes to the dataset when modifying rule sets, and to automate comparison of the rule sets, so that he can compare similarities and differences between logic rule sets. The motivating example highlights the following automating features that would help Jack in extracting logic rule sets. 

\begin{itemize}
    \item Querying a world model to generate the initial set of logic rules requires effective prompting (i.e. what prompt works)
    \item Generating responses using effective strategies that can be interpretable and executable in code
    \item Extracting domain-specific variables for a particular domain
    \item Showing similarities and differences between logic rules for extracted domain-specific variables
    \item Eventually, automating requirements' elicitation process for extracting logic rules
\end{itemize}

\section{RuleFlex} 


As shown in \autoref{fig:proposed_approach}, we propose RuleFlex, an approach designed for implementing and validating rules. RuleFlex has four components: 1) linguistic interface, 2) rule generation engine, 3) dynamic rule modifier, and 4) API generator. Developer describes the problem and specifies configurations using the linguistic interface. The linguistic interface queries a world model to generate the initial set of logic rules through different prompt engineering techniques, and includes the format of responses in prompts so that LLM responses are interpretable and executable. 



The rule generation engine generates rule sets using large language model as a world model, identifies variables and overlap between logic rule sets among domain-specific variables in generated rule sets. An example of domain-specific variables for building software in healthcare includes body temperature, respiratory rate, oxygen saturation, sore throat, and shortness of breath. An example of merge conflict between logic rule sets is `Rule set 1: IF (Body temperature >= 38) THEN RED' versus `Rule set 2: IF (Body temperature >= 37.5) AND (Shortness of breath == NO) THEN RED'. In this example, the rule generation engine compares each logic rule, and classifies the match as wrong threshold, extra condition, and wrong operator. It will then assign a score for each classification and show similarities and differences between rule sets to the developer. 

The dynamic rule modifier allows SMEs and developers to collaboratively modify logic rule sets and validates the generated rules. Finally, the API generator produces API modules and test suite that can be deployed into the production environment. 



\section{Experiment Design}

In this section, we discuss the design of experiments executed for evaluating the proposed approach. This paper focuses specifically on examining the linguistic interface and the rule generation engine. 

\subsection{Prompt Engineering Techniques}

Several prompt engineering techniques that have been studied \cite{zhou2022large, liu2023pre, white2023chatgpt, wei2022chain, zhu2023large, kojima2022large, prystawski2023think, shanahan2023role}. We narrowed the selection to instruction following \cite{kojima2022large, prystawski2023think}, imitation \cite{shanahan2023role}, chain of thought \cite{wei2022chain}, and few-shot \cite{zhu2023large}. Instruction following offers step-by-step guidance \cite{kojima2022large, prystawski2023think}, imitation does role prompting similar to subject-matter expert \cite{shanahan2023role}, chain of thought has reasoning capacity \cite{wei2022chain}, and few-shot prompting has an iterative nature and refines responses based on inputs \cite{zhu2023large}. Out of all prompt engineering techniques, few-shot (i.e. one shot) prompting closely represented the PiMS rule set \cite{logothetis2022pims}. The used few-shot prompt is as follows: 

\textit{You are subject-matter expert (SME). You are helping a software development team to build software for your domain. You do this by providing rule sets that are executable in Python code, using variables to determine actions/outcomes. Rule set formatting: \newline 
"""""" The variables used must be collected by digital systems. For example: \newline
Problem domain: Financial services \newline
Objective: Fraud detection \newline
allowedTransactionAmount = 50000 \newline
transactionType = `Daily' \newline
currency = `USD' \newline
fraudDetected = `NO' \newline
\newline
def fraud\_detection\_rule (transaction\_amount, transaction\_type, transaction\_currency): \newline
\blank{4mm} if (transaction\_amount <= allowedTransactionAmount) and (transaction\_currency != currency): \newline
    \blank{8mm} fraudDetected == `POSSIBLE' \newline
    else: \newline
        \blank{8mm} if (transaction\_amount > allowedTransactionAmount) and (transaction\_type != transactionType): \newline
            \blank{12mm} fraudDetected == `YES' \newline
        \blank{8mm} else: \newline
            \blank{12mm} fraudDetected == `NO' \newline
"""""" \newline
Problem domain: Medical \newline
Objective: Classify COVID-19 patient's health status for developing pandemic intervention monitoring system to help clinicians determine if patient should receive medical attention. The classification must be as follows: GREEN, AMBER, RED.}

\subsection{Large Language Models}

Commercially available large language models such as GPT-3.5 and GPT-4\footnote{https://openai.com/} were selected for the purpose of this study based on their popularity (i.e. under extensive investigation in recent research) \cite{white2023chatgpt, wei2022chain, hou2023large, zhu2023large}. Open source LLMs were excluded from the study as they (i) underperform \cite{zhao2023survey}, (ii) require extensive infrastructure to operate, and (iii) have been trained on smaller datasets \cite{bommasani2021opportunities}. To run experiments, we used playground\footnote{https://platform.openai.com/playground} for GPT-3.5 (i.e. GPT-3.5-Turbo) and GPT-4 with the few-shot prompt. We set temperature to 1, maximum length of response token count to 3000, capability as chat completion, and made one call per response because it is a single prompt.


\subsection{Evaluation Research Questions}

The following evaluation research questions were identified to evaluate our approach:
\begin{itemize}
    \item \textit{RQ1 (Interpretability): How does the extracted world knowledge compare with expert defined knowledge?}
    To answer RQ1, we investigate few-shot prompting to generate rules using the two LLMs (GPT-3.5 and GPT-4) and compare responses in terms of total number of rule sets, and average number of conditions. We measure interpretability in terms of number of rule sets \cite{yang2021learning, zhang2020diverse}, and number of conditions \cite{zhang2020diverse}. An example of condition in a rule `IF (Body temperature >= 38) THEN RED' is `Body temperature >='. A rule set is a combination of rules\footnote{https://github.com/csinva/imodels}. For the number of rule sets, the total is considered across a rule set, extracted from the LLM response. For the number of conditions, average is considered across a ruleset, extracted from the LLM response.


    
    
    \item \textit{RQ2 (Accuracy): Does an LLM defined knowledge system perform as well as human defined?} 
    To answer RQ2, we investigated rules generated by few-shot prompting using two LLMs (GPT-3.5 and GPT-4) and compare responses in terms of overlap between variables in domain-specific variables. In previous studies, similarity between rule sets has been considered as accuracy \cite{zhang2020diverse, perner2013compare}. However, accuracy is impacted by interpretability \cite{nanfack2021global}.

    

    \item \textit{RQ3 (Consistency): How consistent is LLM defined knowledge system?} 
    To answer RQ3, we run experiments 10 times and compare responses in terms of total number of rule sets, and average number of conditions. 
\end{itemize}

\section{Experiment Results}

For interpretability (RQ1), \autoref{tab:total_rules_and_average_conditions} shows the total number of rule sets and average number of conditions across each rule set for few-shot prompting using two different LLMs. However, PiMS had 16 risk rule sets and 25 scoring rule sets, which in total resulted in 41 rule sets. For GPT-3.5 total number of rule sets varied between 3 and 4 rule sets, and for GPT-4 total number of rule sets varied between 2 and 4 rule sets. GPT-3.5 produced an average of 2 to 4 conditions, while GPT-4 showed an average ranging from 2 to 8 conditions.

\begin{table}[htb]
\centering
\begin{tabular}
{p{.15\linewidth}|p{.3\linewidth}|p{.3\linewidth}}
\hline
\textbf{LLM} & \textbf{Total number of rule sets} & \textbf{Average number of conditions} \\
\hline
GPT-3.5 & 3 (3.2) & 3 (3.1) \\
GPT-4 & 3 & 5 (4.5) \\
\hline
\end{tabular}
\caption{Average of total number of rule sets and average of average number of conditions for each rule set, generated by LLMs. The numbers in parentheses represent the precise averages, while the numbers outside are the rounded averages.}
\label{tab:total_rules_and_average_conditions}
\vspace{-5mm}
\end{table}


For accuracy (RQ2), we compared logic rule sets generated from few-shot prompting against the logic rule sets used in PiMS in terms of overlap between domain-specific variables (\autoref{tab:domain_specific_variables}) and comparison of overlapping of the rules that apply to those domain-specific variables (\autoref{tab:accuracy_gpt_pims}). In \autoref{tab:domain_specific_variables}, comorbidity refers to hypertension, lung disease, cardiac disease, immunosuppressed, and diabetes. Both GPT-3.5 and GPT-4 failed to mention domain-specific variables (i.e. symptoms) such as myalgia, diarrhoea, and runny nose, which PiMS had covered. 

\begin{table}[htb]
\centering
\begin{tabular}
{p{.35\linewidth}|p{.15\linewidth}|p{.15\linewidth}|p{.15\linewidth}}
\hline
\textbf{Domain-specific Variable} & \textbf{GPT-3.5} & \textbf{GPT-4} & \textbf{PiMS} \\
\hline
Body Temperature & \checkmark  & \checkmark & \checkmark \\
Shortness of Breath &  & \checkmark & \checkmark \\
Cough & \checkmark  & \checkmark & \checkmark \\
Loss of Taste or Smell &  & \checkmark & \checkmark\\
Sore Throat &  &  & \checkmark \\
Respiratory Rate & \checkmark  &  & \checkmark \\
Fatigue & \checkmark  & \checkmark & \checkmark\\
Oxygen Saturation &  & \checkmark  & \checkmark \\
Heart Rate & \checkmark  & & \checkmark \\
Age &  & \checkmark & \checkmark \\
Comorbidity  &  & \checkmark & \checkmark \\
Gender &  & & \checkmark \\
Myalgia &  & & \checkmark \\
Diarrhoea &  & & \checkmark \\
Runny Nose &  & & \checkmark \\
\hline
\end{tabular}
\caption{Comparison of domain-specific variables between LLMs and PiMS}
\label{tab:domain_specific_variables}
\vspace{-5mm}
\end{table}

\begin{table}[htb]
\centering
\begin{tabular}
{p{.2\linewidth}|p{.15\linewidth}|p{.15\linewidth}|p{.15\linewidth}|p{.15\linewidth}}
\hline
\textbf{Accuracy} & \textbf{Match} & \textbf{Wrong Threshold} & \textbf{Extra Condition} & \textbf{Wrong Operator} \\
\hline
GPT-3.5 vs PiMS & 1 & 5 & 19 & 9 \\
GPT-4 vs PiMS & 7 & 16 & 43 & 18 \\
\hline
\end{tabular}
\caption{Accuracy when comparing LLM generated rules with PiMS rules}
\label{tab:accuracy_gpt_pims}
\vspace{-5mm}
\end{table}


For consistency (RQ3), both GPT-3.5 and GPT-4 responses are not consistent among responses of each time the experiment was run. All experiment results are available here\footnote{https://figshare.com/s/2377e49819fa04703203}.

\section{Related Work}

Recent advances in LLMs have sparked interest in their ability to learn rules and patterns from textual data. However, the previous study relies on labelled datasets, which can be difficult to obtain at the start of new projects due to ethical constraints \cite{lavin2022technology}. LLMs have shown promise in software engineering tasks by exploiting statistical patterns in source code \cite{austin2021program, chen2021evaluating, feng2020codebert}. However, these models are not tailored for discovering new rules, as they rely heavily on similarities to existing code. During requirements elicitation, developers may need to explore rules before any code is written or validate expert-provided rules after system deployment in situations where code completion is insufficient. While prior works have explored bootstrapping requirements \cite{barnett2015bootstrapping}, the generation of logic rules or rule sets for implementation has not been investigated. 

For learning rules, previous research has explored different approaches \cite{mostafaei2022dealing, yang2021learning}. One approach considers stages of preprocessing, rule extraction, and post-processing to handle data difficulties such as outliers. However, evaluation is limited to binary classification on balanced datasets \cite{mostafaei2022dealing}. \citet {yang2021learning} propose rule learning as two subset selection problems, demonstrating gains in scalability and interpretability. However, this is possibly hindered by the rigid two-stage process. Despite methods based on learning ``poor'' weak rules showing to be promising, they are not robust to noisy labels \cite{yang2021learning}.

In healthcare, LLMs have been applied to diagnosis, prognosis, and monitoring, sometimes augmented with visual inputs \cite{balas2023conversational, zhou2023skingpt, liu2023radiology, zhang2023huatuogpt, feng2023large, wang2023chatcad, shyr2023identifying}. However, these models can generate plausible but incorrect responses with high confidence \cite{balas2023conversational, liu2023radiology} without asking clarifying questions \cite{balas2023conversational, zhou2023skingpt}. LLMs also pose risks around privacy and lack empathy \cite{zhou2023skingpt, liu2023radiology, zhang2023huatuogpt}. Prior research \cite{balas2023conversational, zhou2023skingpt, liu2023radiology, zhang2023huatuogpt, feng2023large} has extended the existing LLMs to explore their effectiveness in diagnosis, however, we utilize LLMs without any modifications. Despite promising results on assessing responses, no studies have deployed LLMs in real-world clinical settings \cite{rao2023evaluating, zhou2023skingpt, liu2023radiology, singhal2023large, singhal2023towards, hiesinger2023almanac, liu2023using}. LLMs show promise for learning rules and patterns from text; however, most applications focus on well-defined tasks over curated data. Their effectiveness at discovering new rules in real-world settings in different domains such as healthcare remains relatively unexplored.

\section{Conclusion and Future Work}

In software systems, maintaining logic rules/rule sets has similar challenges of maintaining code base. However, logic rule sets are designed for quick turnaround in software engineering. Therefore, identifying similarities and differences between logic rule sets is vital. LLMs have an understanding based on internet scale datasets, which includes a model for every domain. Therefore, they are best suited for software engineering tasks prior to human review. The proposed approach demonstrates the potential for learning rules from natural language. Further research is required to expand its applicability. The effectiveness of LLMs in generating intricate logic rules relies on domain-specific information, which may be beyond the awareness of LLMs. Adapting the approach for image datasets is one area for future work, as it currently handles textual and logic rules. Exploring additional prompt engineering techniques could also improve performance. Furthermore, we have considered only four prompt engineering techniques where other prompt engineering techniques such as graph of thoughts \cite{besta2023graph}, and a combination of prompt engineering techniques \cite{chen2023unleashing} can improve performance (future exploration). 

In addition, our evaluation was on one domain-specific dataset, limiting the generalisation of our findings. Future work includes evaluating our approach to other domain-specific datasets to improve generalisability such as finance, law, and other applied fields. We have considered only two dimensions, such as interpretability and accuracy. Therefore, we have not considered other factors such as trustworthy AI, fairness, and robustness. The field of LLMs is rapidly evolving with the introduction of new models. In the current study, we have considered only two LLMs. There are other types of LLMs such as LLaMA, Alpaca, Vicuna, Falcon, MPT, and LLaMA2, and domain-specific LLMs such as Med-PaLM\footnote{https://sites.research.google/med-palm/} which may have improved rule learning in those applied contexts. Overall, we can build on this work by broadening the data types, prompt engineering techniques, evaluation metrics, architectures, and experimental designs. 



\bibliographystyle{ACM-Reference-Format}
\bibliography{references}


\end{document}